\documentstyle[aps]{revtex}

\begin{document}
\author{A. Zh. Muradyan$^{1,2}$, V. A. Poghosyan$^1$}
\address{$^1$Department of Physics, Yerevan State University, 1 Alex Manukian, Yerevan%
\\
Armenia\\
$^2$Engineering Center of Armenian National Academy of Sciences, Ashtarak-2,%
\\
378410 Armenia;\\
E-mail:muradyan@ec.sci.am; yndanfiz@ysu.am.}
\title{Optical Transition and Momentum Transfer in Atomic Wave Packets}
\maketitle

\begin{abstract}
It is shown that the population Rabi-floppings in a lossless two-level atom,
interacting with a monochromatic electromagnetic field, in general are
convergent in time. The well-known continuous floppings take place because
the restricted choosing of initial conditions, that is when the atom
initially is chosen on ground or excited level before the interaction,
simultaneously having a definite value of momentum there. The convergence of
Rabi-floppings in atomic wave-packet-states is a direct consequence of
Doppler effect on optical transition rates (Rabi-frequencies): it gradually
leads to ''irregular'' chaotic-type distributions of momentum in ground and
excited energy levels, smearing the amplitudes of Rabi-floppings. Conjointly
with Rabi-floppings, the coherent accumulation of momentum on each internal
energy level monotonically diminishes too.
\end{abstract}

\section{Introduction}

It is well known that due to interaction with a plane travelling wave the
atomic momentum may be changed in limits of one photon momentum $\hbar k$.
This limitation follows from momentum conservation law, and, concerns to
total atomic momentum. As to momentum per each internal energy level, as was
shown recently \cite{Muradyan}, its change may be great and even much-more
surpass the coupling photon's momentum $\hbar k.$ It is the case, when the
atom initially is in superposition state of lower and upper energy levels
with some (different in general) momentum distributions there. In other
words, optical transition between atomic wave packet-states is accompanied
by large-scale coherent accumulation of momentum in internal energy levels
(CAMEL).

This phenomenon, which, as we hope, will have far-going consequences for
atomic and molecular physics, is presented in \cite{Muradyan} in the form as
simple as possible. In particular, the operator of kinetic energy of atomic
transitional motion was not included into the Hamiltonian of the atom-field
system. Nevertheless this operator has not only quantitative, but also
qualitative contribution into the picture of interaction. For instance, such
an important phenomenon as is the Doppler-shift of frequencies, is
introduced into the theory (in laboratory frame) by means of mentioned
operator. Therefor, in all cases, the more logical theory of atom-field
interaction, concerning the atomic wave packet-states, should contain the
atomic kinetic energy operator unquestionably. Just this is done in
presented paper, that is here we consider an optical transition in the
two-level atom, which has been prepared in general quantum-mechanical
transitional states for lower and upper internal energy levels. The behavior
of level population and momentum transfer between energy levels is
considered in details. It is shown, that the CAMEL-phenomenon, which
exhibits regular-periodic behavior in time when the kinetic energy operator
is not taken into account, really has a damping-periodic behavior and it is
due to influence of the Doppler-shift of frequencies on the rate of optical
transitions (Rabi-frequencies). Simultaneously a strictly important result
has been obtained for atomic internal levels populations (population
amplitudes), according to which the Rabi-floppings of populations for
wave-packet atomic states have a damping behavior in general. It is
worthwhile to remind that the well-known continuous periodic behavior takes
place for ''ordinary'', fully unexcited or fully excited pure initial
states, that is when the time evolution of populations begins from only one
populated internal energy level, and, in addition, this populated state has
definite value of momentum.

The quantum-mechanical behavior of a two-level atom in the near-resonant,
plane-wave monochromatic radiation, taking into account the atomic kinetic
energy operator, was considered earlier many times, in most close-staging to
our, in papers \cite{Pusep}, \cite{Zheng}. In \cite{Pusep} the atom
initially is on one energy level and the analysis is limited by narrow
momentum distributions and short times of interaction. As a consequence, a
splitting of extra-narrow wave-packet into two subpackets has been created
due to interaction. In paper \cite{Zheng} the authors restricted the
analysis by the definite momentum and one energy level population case. As a
consequence, only the continuous periodic behavior, taking into account the
energy level splittings due to photon momentum exchange, has been obtained
there for energy level's populations.

Taking into account the results of this paper, the following may be stated
about the role of initial conditions. For more general quantum-mechanical
initial states, including the atomic transitional states, a) the
Rabi-floppings have damping in time character and b) the optical transitions
are accompanied by coherent accumulation of momentum on the internal energy
levels (CAMEL). Choosing population only on the one internal energy level
annihilates the possibility of CAMEL-phenomenon. If, in addition, the
populated state has only a definite value of momentum, then the damping
behavior of Rabi-floppings disappears too. Only for simplest initial
conditions the behavior of interaction obtains continuous periodic nature
with exchange of momentum between internal energy levels restricted by one
photon momentum $\hbar k$.

This paper is organized as follows. The Schrodinger equation and the
stationary solutions in momentum representation are presented in Sec. 2. The
time evolution of level populations and the transfer of momentum and kinetic
energy between internal energy levels are examined in Sec. 3 and 4
respectively. The paper is finished by short Conclusions.

\section{Rotating-wave approximation stationary solutions of Schrodinger
equation in momentum representation.}

Let us consider the near-resonant interaction of a two-level atom with a
plane-wave radiation field \cite{Allen}. The Hamiltonian of this system is
well known and, taking into account the translational motion of atom, can be
presented in the form

\begin{equation}
\hat{H}=\frac{\hat{p}^2}{2M}+\frac{\hbar \omega _0}2\hat{\sigma}_3+\hat{V},
\label{1}
\end{equation}
where $M$ and $\omega _0$ denote respectively the atomic mass and optical
transition frequency, $\hat{\sigma}_3$ is quasispin (Pauli) operator. Second
term presents the free atom with $-\frac{\hbar \omega _0}2$ and $\frac{\hbar
\omega _0}2$ energies in lower and upper energy levels. Last term, $\hat{V}$%
, presents the interaction of atom with external travelling-wave field, and
can be written in dipole approximation as

\begin{equation}
\hat{V}=-\hat{d}E(t,z),  \label{2}
\end{equation}
where $\hat{d}$ is dipole moment operator for optical transition and the
intensity of the plane travelling wave we'll present in the form 
\begin{equation}
E(t,z)=E_{0}\exp (ikz-i\omega t)+c.c.  \label{3}
\end{equation}
with $E_{0}${\bf \ }is constant, $\omega $ and $k=\omega /c$ represent the
wave frequency and the wave number. Polarization effects aren't included
into the field of investigation. Such approach is valid, as is well known,
for purely linear or circular polarizations of the wave.

Denoting by $\varphi _{g}(\overrightarrow{\rho },t)$ and $\varphi _{e}(%
\overrightarrow{\rho },t)$ the wave functions of ground ($g$) and excited ($%
e $) energy levels ($\overrightarrow{\rho }$ is atomic internal coordinate,
i.e. the radius vector of optical electron relative to atomic
center-of-mass), the wave function of interacting atom may be written in the
following form 
\begin{equation}
\Psi (\overrightarrow{\rho },z,t)=A(z,t)\text{ }\varphi _{g}(\overrightarrow{%
\rho },t)+B(z,t)\text{ }\varphi _{e}(\overrightarrow{\rho },t),  \label{4}
\end{equation}
where $A(z,t)$ represents the atomic probability amplitude to be on lower
level and have a space-coordinate z at the time moment t; and the other
coefficient $B(z,t)$ represents the same for upper level atom. Note, that
the coordinate $z$ in (\ref{4}) represents the atomic center-of-mass
position in wave direction; and hence the plane wave (\ref{3}) includes only
this single variable $z$, the case can be considered as a question of one
dimension.

In this paper our attention will be focused onto the time evolution of
atomic momentum distributions on lower and upper internal energy levels and
their physical consequences. So hereafter it is worthwhile to deal with
atomic amplitudes in the momentum representation. Moreover, just in momentum
representation the eigenvalue problem for the system under consideration has
analytic solutions \cite{Pusep}, \cite{Zheng}.

Expanding $A(z,t)$ and $B(z,t)$ amplitudes into momentum space on the basis
of definite momentum states 
\begin{equation}
\chi (p)=\frac 1{\sqrt{2\pi \hbar }}\exp (\frac i\hbar p\text{ }z),
\label{5}
\end{equation}

that is

\begin{equation}
A(z,t)=\int a(p,t)\chi (p)dp;\ B(z,t)=\int b(p,t)\chi (p)dp,  \label{6}
\end{equation}
we substitute all related quantities (\ref{1})-(\ref{6}) into the
Schrodinger equation

\begin{equation}
i\hbar \frac{\partial \Psi }{\partial t}=\hat{H}\Psi  \label{7}
\end{equation}
After standard transformations we arrive for seeking amplitudes $a(p,t)$ and 
$b(p,t)$ to

\begin{eqnarray}
i\hbar \frac{da(p,t)}{dt} &=&\left( \frac{p^{2}}{2M}+\frac{\hbar \omega _{0}%
}{2}\right) a(p,t)-\frac{\hbar \Omega }{2}e^{-i\omega t}b(p-\hbar k,t),
\label{8} \\
i\hbar \frac{db(p,t)}{dt} &=&\left( \frac{p^{2}}{2M}-\frac{\hbar \omega _{0}%
}{2}\right) b(p,t)-\frac{\hbar \Omega }{2}e^{i\omega t}a(p+\hbar k,t),
\label{9}
\end{eqnarray}
where $\Omega =2dE_{0}/\hbar $ is the parameter of induced transitions and
commonly referred to as Rabi frequency \cite{hing}. This system fully
coincide in form with the system of equations (8), (9) in \cite{Zheng}, and
its stationary form should be coincide with the system (5) in \cite{Pusep}.

General solution of (\ref{8}) and (\ref{9}) is

\begin{eqnarray}
a(p,t) &=&-\left( \frac{\alpha (p)-\beta (p)}{2\beta (p)}a(p,0)+\frac{\Omega 
}{2\beta (p)}b(p+\hbar k,0)\right) e^{-i\omega _{g}(p)t}+  \nonumber \\
&&+\left( \frac{\alpha (p)+\beta (p)}{2\beta (p)}a(p,0)+\frac{\Omega }{%
2\beta (p)}b(p+\hbar k,0)\right) e^{-i\omega _{g}^{^{\prime }}(p)t}\text{ \ }%
,  \label{10} \\
b(p+\hbar k,t) &=&\left( -\frac{\Omega }{2\beta (p)}a(p,0)+\frac{\alpha
(p)+\beta (p)}{2\beta (p)}b(p+\hbar k,0)\right) e^{-i\omega _{e}(p)t}+ 
\nonumber \\
&&+\left( \frac{\Omega }{2\beta (p)}a(p,0)-\frac{\alpha (p)-\beta (p)}{%
2\beta (p)}b(p+\hbar k,0)\right) e^{-i\omega _{e}^{^{\prime }}(p)t}\text{ \ }%
.  \label{11}
\end{eqnarray}

Here 
\begin{equation}
\alpha (p)=\frac{\hbar k^{2}}{2M}+\frac{pk}{M}+\Delta
\end{equation}
and may be viewed as a generalized detuning, which involves the field-atom
detuning $\Delta =\omega _{0}-\omega $, Doppler- and recoil-shifts $\frac{pk%
}{M}$ and $\frac{\hbar k^{2}}{2M}$ respectively. It really represents the
usual field-atom detuning viewed from atomic center-of-mass frame of
reference. The second term 
\begin{equation}
\beta (p)=\sqrt{\left( \frac{\hbar k^{2}}{2M}+\frac{pk}{M}+\Delta \right)
^{2}+\Omega ^{2}}  \label{13}
\end{equation}
and represents merely the so called generalized Rabi frequency, including
the generalized detuning $\alpha (p)$ instead of common frequency detuning $%
\Delta $. Primed and nonprimed frequencies in exponents are 
\begin{eqnarray}
\omega _{g,e}^{^{\prime }}(p) &=&\frac{1}{2\hbar }\left( \frac{p^{2}}{2M}+%
\frac{(p+\hbar k)^{2}}{2m}\mp \hbar \omega \right) -\frac{\beta (p)}{2},
\label{14} \\
\omega _{g,e}(p) &=&\frac{1}{2\hbar }\left( \frac{p^{2}}{2M}+\frac{(p+\hbar
k)^{2}}{2m}\mp \hbar \omega \right) +\frac{\beta (p)}{2},  \label{15}
\end{eqnarray}
and represent the energies of system quasistationary states,
double-splitted, as in familiar theory, in both excited and ground levels.
The size of splittry is $\omega _{g}-\omega _{g}^{^{\prime }}=\omega
_{e}-\omega _{e}^{^{\prime }}=\beta (p)$. Note, that plugging in (\ref{10}),
(\ref{11}) $b(p,0)=0,$ $a(p,0)=\delta (p-p_{0})$, we arrive to the case,
analyzed in paper \cite{Zheng}.

\section{Population and momentum per internal ground and excited energy
levels}

Let's now proceed to calculation of such physical quantities, as population
and mean momentum in each atomic internal energy level, and to their
distributions in momentum space. Time evolution of population distributions
has been determined and is presented by (\ref{10}) and (\ref{11}). Main
peculiarity of these formulas is their $p$-dependence due to Doppler effect,
which will play the key role in further presenting results. It, first of
all, disturbs the population distribution in momentum space and gradually
transforms the initial smooth distribution into the modulated, chaotic-like
one. A behavior of time-evolution is illustrated in Fig. \ref{Fig1v}. Curve $%
a$ represents the initial distribution, curve $b$ - after 4 Rabi floppings
(for central range of distribution) and curve $c$ - after 12 Rabi floppings.
For simplicity the excited level was assumed to be initially empty, and the
preserving symmetry about $p=0$ value is conditioned by this assumption. In
general, when both energy levels are populated, even symmetric with respect
to some values of momentum, any symmetry in distribution is being lost
rapidly.

Total population of internal energy level is 
\begin{equation}
n_{g}=\int \left| a(p,t)\right| ^{2}dp  \label{16}
\end{equation}
for ground energy level, and is 
\begin{equation}
n_{e}=\int \left| b(p,t)\right| ^{2}dp  \label{17}
\end{equation}
for excited energy level. The typical form of time evolution for these
populations is presented in Fig. \ref{Fig2v}. As is seen, the Rabi-floppings
(oscillations) are gradually flatted due to redistributions of momentum
states for interaction time. So, the momentum-dependence of probabilities
for optical transitions (Rabi-floppings), arising due to Doppler-shift of
frequencies, leads in general to damping in population oscillations and to
establishment of definite-value populations in internal energy levels
without any mechanism of relaxation.

The atomic momentum

\begin{equation}
\left\langle p\right\rangle =\int \Psi ^{\ast }\widehat{p}\Psi d%
\overrightarrow{\rho }dz,  \label{18}
\end{equation}
after elementary substitution of general expression (\ref{4}) and respective
standard transformations can be expressed as a sum of two terms \cite
{Muradyan}, 
\begin{equation}
\left\langle p\right\rangle =\left\langle p\right\rangle _{g}+\left\langle
p\right\rangle _{e}  \label{19}
\end{equation}
first of which represents the amount of contribution of ground level states
into the atomic momentum and is presented in general as 
\begin{equation}
\left\langle p\right\rangle _{g}=\int \left| a(p,t)\right| ^{2}pdp,
\label{20}
\end{equation}
the second term has the same sense for excited level and is 
\begin{equation}
\left\langle p\right\rangle _{e}=\int \left| b(p,t)\right| ^{2}pdp
\label{21}
\end{equation}

It should be mentioned that these quantities, besides being the ingredients
of total atomic momentum, in accordance with first principles of quantum
mechanics, have own physical meaning and are measurable quantities \cite
{Muradyan}.

Typical form of time-evolutions, obtained by means of numerical
calculations, is illustrated in Fig. \ref{Fig3v} and \ref{Fig4v},
respectively for ground and excited levels. Solid curves represent the case,
when only one energy level (ground), is initially populated, and dashed
curves - when both levels are populated. Herewith, the same forms of
momentum distributions are chosen in the last case and these distributions
are shifted with each other by $p_{0}\gg \hbar k$. In both cases, particular 
$A$, and more general $B$, the oscillatory behavior is being depressed.

To carry out the behavior of CAMEL-phenomenon from the behavior of mean
momentums $\left\langle p\right\rangle _{g}$ and $\left\langle
p\right\rangle _{e}$, it is necessary to pick out from these momentums the
parts, conditioned by population evolution. To this end a pair of new
momentums can be introduced into the theory \cite{Muradyan},

\begin{equation}
p_{g}=\left\langle p\right\rangle _{g}/n_{g}\text{ and }p_{e}=\left\langle
p\right\rangle _{e}/n_{e}  \label{22}
\end{equation}
time evolution of which should be solely conditioned by redistributions in
wave-packet momentum states, referred to as coherent accumulation of
momentum on internal energy level \cite{Muradyan}; for familiar optical
transitions with continuous periodic Rabi-floppings $p_{g}$ and $p_{e}$ are
constant in the course of time and emerge as mean values of normalized
momentum distribution per each energy level. It is readily also verified
that for familiar optical transitions, beginning from fully populated energy
levels, these momentums can be varied in limits of one photon momentum $%
\hbar k$. In the contrary, in general case of initially populated states,
there are essential redistributions in each energy level momentum states
(see, for instance, Figs. \ref{Fig3v} and \ref{Fig4v}) and $p_{g}$ and $%
p_{e} $ characteristic momentums are being changed in essentially exceeding
the photon momentum limits. The graphs of normalized momentums $p_{g}$ and $%
p_{e} $, corresponding to the familiar case, wave-packet case and most
general case are presented in Figs. \ref{Fig5v}-\ref{Fig7v}. As one sees
from Fig. \ref{Fig5v}, the normalized momentums $p_{g}$ and $p_{e}$ in
familiar case are constant during the interaction. The reason is the fact,
that the mean level-momentums $\left\langle p\right\rangle _{g}$ and $%
\left\langle p\right\rangle _{e}$ strictly follow to corresponding
level-populations $n_{g}$ and $n_{e}$ in time, have their form of
time-dependence, leading trivially to constant values for $%
p_{g}=\left\langle p\right\rangle _{g}/n_{g}$ and $p_{e}=\left\langle
p\right\rangle _{e}/n_{e}$. For wave-packet initial states, the normalized
momentums $p_{g}$ and $p_{e}$ aren't constant in time (Fig. \ref{Fig6v} and
Fig. \ref{Fig7v}) but, nevertheless, if only one of initial internal energy
levels is populated, the variations of momentums are restricted by one
photon momentum (Fig. \ref{Fig6v}). The CAMEL-phenomenon appears for both
populated internal energy levels (Fig. \ref{Fig7v}). For initial interval of
interaction time it has essential oscillations, which later, in full analogy
with Rabi-floppings, is depressed. It is worthwhile, probably, to note that
the depressing concerns to oscillations only, but not to CAMEL-phenomenon; $%
p_{g}$ and $p_{e}$ go to definite values, different from their initial
values. The reason of such ''saturation'' is the fact, that the momentum
distributions become so much chaotic-type, that the further ''chaotization''
has much more slow influence on the system evolution, in particular, on the
CAMEL.

\section{Translational energy per internal ground and excited energy levels.}

An analogous set of calculations we have made also for the kinetic energy of
atom. It can be readily verified by standard calculations, that the (\ref{19}%
)-type splitting is true also for kinetic energy: 
\begin{equation}
\left\langle E_{kin}\right\rangle =\left\langle E_{kin}\right\rangle
_{g}+\left\langle E_{kin}\right\rangle _{e},  \label{23}
\end{equation}
with 
\begin{eqnarray}
\left\langle E_{kin}\right\rangle _{g} &=&\int \left| a(p,t)\right| ^{2}%
\frac{p^{2}}{2M}dp,  \label{24} \\
\left\langle E_{kin}\right\rangle _{e} &=&\int \left| b(p,t)\right| ^{2}%
\frac{p^{2}}{2M}dp,  \label{25}
\end{eqnarray}
presenting the part contribution of ground and excited internal energy
levels into the atomic kinetic energy. Time-behavior of (\ref{24}) and (\ref
{25}) for one-level initial population $(a(p,0)\neq 0,b(p,0)=0)$ case
exhibits, of course, continuously flopping behavior for familiar one-state
population case (Fig. \ref{Fig8vab}a) and oscillatory saturating behavior
for one-level wave-packet case (Fig. \ref{Fig9vab}a). The general, two-level
population case with different wave-packet distributions, is presented in
Fig. \ref{Fig10vab}a. For comparison we present there also the total kinetic
energy of atom (solid lines in Figs. \ref{Fig9vab}-\ref{Fig10vab}). As seen
from Figs. \ref{Fig9vab} and \ref{Fig10vab}, total energy has some
fluctuations after instant switching on of the interaction, which gradually
disappears. That is because of Heisenberg principle of indeterminacy.

The behavior of normalized quantities, 
\begin{eqnarray}
E_{kin,g} &=&\left\langle E_{kin}\right\rangle _{g}/n_{g},  \label{26} \\
E_{kin,e} &=&\left\langle E_{kin}\right\rangle _{e}/n_{e},  \label{27}
\end{eqnarray}
are presented in Figs. \ref{Fig8vab}b-\ref{Fig10vab}b. The role of initial
state preparation is obvious for kinetic energy too and is totally similar
to the case of momentum behavior.

Nevertheless, there is a very important difference between the behaviors of
momentum and energy, if we discriminate internal and external degrees of
atom for the interaction. The coupling photonic momentum can only transfer
(or is obtained from) external translational degrees of atom, while the
photonic energy deals with both internal (and it is the main part) and
external degrees of freedom. As a consequence, a logically correct scale, as
is the photon momentum for atomic translational motion momentum, absent for
atomic translational motion energy. That is why the statement of the
question about CAMEL-type phenomenon in the translational energy space
quantitatively looks to be problematic, at least hitherto.

\section{Conclusions}

The detailed analysis of a two-level atom in the field of near-resonant
monochromatic radiation shows, that in more general quantum-mechanical
states with some distributions in momentum space (wave-packet states) the
Rabi-floppings between internal energy levels monotonically decrease for
long times of interaction. The reason of decreasing is momentum dependence
of optical transition probabilities, or, in more usual in laser physics
terminology, due to dependence of Rabi-floppings on the Doppler-shift of
frequencies.

\begin{center}
\begin{figure}[tbp]
\caption{Population distribution in momentum space for ground state.}
\label{Fig1v}
\end{figure}

\begin{figure}[tbp]
\caption{Internal levels populations for one-level wave-packet initial case.
Hereafter X-axes is measured in $\protect\tau =\frac{\hbar k^{2}}{2M}t$
units.}
\label{Fig2v}
\end{figure}

\begin{figure}[tbp]
\caption{Mean momentas of internal ground energy level. $A$ - for familiar
one-state wave-packet case, and $B$ - for general superpositional case.}
\label{Fig3v}
\end{figure}
\begin{figure}[tbp]
\caption{Mean momentas of internal excited energy level. $A$ - for familiar
one-state wave-packet case, and $B$ - for general superpositional case.}
\label{Fig4v}
\end{figure}
\begin{figure}[tbp]
\caption{Normalized momentums $p_{g}$ and $p_{e}$ for familiar one-state
case with definite momentum.}
\label{Fig5v}
\end{figure}
\begin{figure}[tbp]
\caption{Normalized momentums $p_{g}$ and $p_{e}$ for one-state wave-packet
case.}
\label{Fig6v}
\end{figure}
\begin{figure}[tbp]
\caption{Normalized momentums $p_{g}$ and $p_{e}$ for general
superpositional case.}
\label{Fig7v}
\end{figure}
\begin{figure}[tbp]
\caption{a) Internal levels energies for familiar one-level state with
definite momentum. b) Internal levels normalized energies for one-level
state with definite momentum.}
\label{Fig8vab}
\end{figure}
\begin{figure}[tbp]
\caption{a) Total energy, and internal levels energies for one-state
wave-packet initial case. b) Total energy, and internal levels normalized
energies for one-state wave-packet initial case.}
\label{Fig9vab}
\end{figure}
\begin{figure}[tbp]
\caption{a) Total energy and internal levels energies for general
superpositional case. b) Total energy and internal levels normalized
energies for general superpositional case.}
\label{Fig10vab}
\end{figure}
\end{center}

\end{document}